\documentclass[sigconf,screen,nonacm]{acmart}

\renewcommand\footnotetextcopyrightpermission[1]{} 
\AtBeginDocument{%
  }

\usepackage{CJKutf8}
\usepackage{listings}
\usepackage{xcolor}
\usepackage{siunitx} 

\definecolor{codegreen}{rgb}{0,0.6,0}
\definecolor{codegray}{rgb}{0.5,0.5,0.5}
\definecolor{codepurple}{rgb}{0.58,0,0.82}
\definecolor{backcolour}{rgb}{0.95,0.95,0.92}

\lstdefinestyle{mystyle}{
    backgroundcolor=\color{backcolour},   
    commentstyle=\color{codegreen},
    keywordstyle=\color{magenta},
    numberstyle=\tiny\color{codegray},
    stringstyle=\color{codepurple},
    basicstyle=\ttfamily\footnotesize,
    breakatwhitespace=false,         
    breaklines=true,                 
    captionpos=b,                    
    keepspaces=true,                 
    numbers=left,                    
    numbersep=5pt,                  
    showspaces=false,                
    showstringspaces=false,
    showtabs=false,                  
    tabsize=2
}

\begin{document}

\title[Exploring the Alignment of Perceived and Measured Sleep Quality with Working Memory]{Exploring the Alignment of Perceived and Measured Sleep Quality with Working Memory using Consumer Wearables}

\author{Peter Neigel}
\email{sw23824q@st.omu.ac.jp}
\orcid{0000-0001-6336-2138}
\affiliation{%
  \institution{Osaka Metropolitan University}
  \city{Sakai}
  \state{Osaka}
  \country{Japan}
}
\affiliation{%
  \institution{RPTU Kaiserslautern-Landau}
  \city{Kaiserslautern}
  \state{Rheinland-Pfalz}
  \country{Germany}
  \postcode{67663}
}

\author{David Antony Selby}
\email{david.selby@dfki.de}
\orcid{0000-0001-8026-5663}
\affiliation{%
  \institution{German Research Center for Artificial Intelligence}
  \city{Kaiserslautern}
  \state{Rheinland-Pfalz}
  \country{Germany}
}

\author{Shota Arai}
\orcid{0009-0009-7737-4952}
\email{sd24111e@st.omu.ac.jp}
\affiliation{%
  \institution{Osaka Metropolitan University}
  \city{Sakai}
  \state{Osaka}
  \country{Japan}
}
\author{Benjamin Tag}
\orcid{0000-0002-7831-2632}
\email{benjamin.tag@unsw.edu.au}
\affiliation{%
  \institution{University of New South Wales}
  \city{Sydney}
  \country{Australia}
}
\author{Niels van Berkel}
\orcid{0000-0001-5106-7692}
\email{nielsvanberkel@cs.aau.dk}
\affiliation{%
  \institution{Aalborg University}
  \city{Aalborg}
  \country{Denmark}
}

\author{Sebastian Vollmer}
\orcid{0000-0003-2831-1401
}
\email{sebastian.vollmer@dfki.de}
\affiliation{%
  \institution{German Research Center for Artificial Intelligence\\RPTU Kaiserslautern-Landau}
  \city{Kaiserslautern}
  \state{Rheinland-Pfalz}
  \country{Germany}
}

\author{Andrew Vargo}
\orcid{0000-0001-6605-0113}
\email{awv@omu.ac.jp}
\affiliation{%
  \institution{Osaka Metropolitan University}
  \city{Sakai}
  \state{Osaka}
  \country{Japan}
}

\author{Koichi Kise}
\orcid{0000-0001-5779-6968}
\email{kise@omu.ac.jp}
\affiliation{%
  \institution{Osaka Metropolitan University}
  \city{Sakai}
  \state{Osaka}
  \country{Japan}
}

\renewcommand{\shortauthors}{Neigel et al.}

\begin{abstract}

Wearable devices offer detailed sleep-tracking data. However, whether this information enhances our understanding of sleep or simply quantifies already-known patterns remains unclear. 
This work explores the relationship between subjective sleep self-assessments and sensor data from an Oura ring over 4--8 weeks in-the-wild. 29 participants rated their sleep quality daily compared to the previous night and completed a working memory task. Our findings reveal that differences in REM sleep, nocturnal heart rate, \textit{N}-Back scores, and bedtimes highly predict sleep self-assessment in significance and effect size. For \textit{N}-Back performance, REM sleep duration, prior night's REM sleep, and sleep self-assessment are the strongest predictors. We demonstrate that self-report sensitivity towards sleep markers differs among participants. We identify three groups, highlighting that sleep trackers provide more information gain for some users than others. Additionally, we make all experiment data publicly available.



\end{abstract}



\maketitle

\thispagestyle{empty}

\section{Introduction}

Understanding and improving sleep quality is a significant concern in clinical and everyday contexts. Sleep is a fundamental aspect of health, influencing performance~\cite{bonnet1989effect} and cognition~\cite{walker2017we}, mood~\cite{Scott_sleep_mood}, and overall quality of life~\cite{bonnet1989effect}. Insufficient sleep has been linked to decreased performance, impaired concentration~\cite{pilcher1997sleep}, and higher levels of self-reported pain~\cite{affleck1996sequential,moldofsky2001sleep}. Therefore, assessing and enhancing sleep quality is essential for optimizing daily functioning and overall life satisfaction. Traditional methods for assessing sleep often rely on subjective self-reports, which can be influenced by biases such as recall bias and response shift, i.e. a change in how individuals evaluate or report their experiences, often due to a shift in their internal standards, values, or understanding of the concept being measured. These biases complicate the accurate measurement of sleep quality and its impact on daily function. The increasing availability of wearable technology has led to a growing interest in whether these devices can enhance our understanding of sleep or quantify known sleep states. While the manufacturers of wearable sleep-tracking devices, such as the Oura Ring, SoXAI, and others, can assert the accuracy of their devices for tracking sleep stages (such as light sleep, deep sleep, and REM sleep), the relationship between the data produced and the user remains unclear. Previous research has shown that wearable health and sleep-tracking devices are often discarded by users~\cite{Gouveia2015Activitytrackers, neigel_UsingWearablesUnobtrusively_2024}, and even when users continue to wear the device, they may fail to initiate any real change in the user's life~\cite{r.nolasco_PerceptionRealityHow_2023}. This shows a potential disconnect between the information provided by the sleep tracker and the ability and motivation of users to understand and internalize the meaning behind the data and act thereon. In other words, we are asking if wearable trackers provide insights beyond what individuals report subjectively or if there are consistent mismatches between reported data and user experience. Understanding this complex relationship is essential to better design devices and interfaces to bridge this gap between sleep-tracking technologies and users. 

Despite dedicating about a third of our adult lives to sleep, many of us experience poor sleep quality~\cite{yi2013sleep}. While standard sleep hygiene practices are well-known, adherence to these recommendations remains low among the general public~\cite{felix2017college}. Accurate sleep assessment is crucial for diagnosing and managing sleep disorders, optimizing performance, and improving overall health. Related research indicates that merely having sleep data may not be enough for users to understand or make practical changes~\cite{arroyo_ImplementationBehaviorChange_2022}. In fact, users may fail to actuate the desired change in sleep behavior, even if they believe they have made such a change~\cite{r.nolasco_PerceptionRealityHow_2023}. Providing additional insights that enhance our understanding of sleep beyond subjective reports could significantly impact how sleep health is monitored and managed. Thus, understanding where self-evaluation and the sleep measurements of wearables meet and diverge is essential for consumers and healthcare providers.

This paper explores the potential gap between subjective sleep assessments and objective data provided by wearable trackers. Our 29 participants collected 1158 consecutive day-pairs of data over 2.5 months. To collect data daily we use the Oura ring\footnote{\url{https://ouraring.com/}}, which has shown good accuracy in comparison to research-grade equipment~\cite{mehrabadi_sleep_2020, cao_AccuracyAssessmentOura_2022a, svensson_ValidityReliabilityOura_2024, chee_MultiNightValidationSleep_2021}, although it possibly underestimates light sleep duration~\cite{kainec_EvaluatingAccuracyFive_2024}. We examine the relationship between daily subjective sleep assessments and sensor-based measures such as heart rate (HR), heart rate variability (HRV) and sleep stages. This research seeks to determine if wearables offer genuine informational benefits to users or if they merely corroborate self-reported sleep quality. We compare relative subjective sleep assessments with objective data. While previous studies have explored the accuracy of sleep trackers~\cite{kainec_evaluating_2024}, our work specifically investigates how well subjective assessments of sleep quality correlate with various objective sleep metrics. Additionally, by including cognitive performance measures, we provide a valuable comparison for linking sleep staging data and self-assessment to real-world performance on a working-memory task that is known to be strongly linked to quality of sleep~\cite{arai_--wild_2024,kuriyama_sleep_2008,peng_effect_2020}

Our findings indicate that differences in REM sleep duration, average nocturnal heart rate, bedtimes, and scores in the working memory test are partially predictive of self-assessed sleep quality. Specifically, REM sleep duration and the prior night
REM sleep was the strongest predictor of cognitive performance in working memory tasks. We identify three groups with varying responses in their self-reports towards the sleep markers, suggesting that while some individuals may benefit from the additional information provided by wearables, others may experience limited information gain from the produced data. 

This paper contributes to the ongoing debate on the utility of wearable sleep trackers by systematically comparing subjective sleep assessments with objective data from the Oura ring.

\subsection{Contributions}

\begin{itemize}
    \item We systematically compare subjective sleep assessments with objective data from the Oura ring, providing insights into the relationship between self-reported sleep quality and wearable sleep metrics.
    
    \item Our analysis identifies key predictors of self-assessed sleep quality, including REM sleep duration, average nocturnal heart rate, and deep sleep metrics, highlighting their significance and effect sizes.
        
    \item By incorporating cognitive performance through an \textit{N}-back task, we offer a comprehensive view of the impact of sleep quality on daily functioning, illustrating the interplay between sleep metrics and cognitive outcomes.

    \item We demonstrate that wearable sleep trackers provide informational gain, the extent of which varies across groups of users, revealing variability in the accuracy of subjective sleep assessments. Our findings contribute to the understanding of what insights wearable sleep trackers offer beyond self-reports, informing both research and practical applications in personal and clinical sleep monitoring.

    \item We make our dataset, including the sleep data, self-reports, and working memory test results, publicly available at [URL anonymized for review; please see supplemental material].
\end{itemize}

The results of our work have broader implications for how we understand and utilize sleep-tracking technology. For individuals who derive added insights from wearables, these tools could enhance self-awareness and improve sleep management. However, for others, the added value of such devices may be limited. 
A better understanding of what constitutes insights for distinct user groups can inform the design of future self-tracking technology that more closely aligns with user needs.

\section{Related Work}

\subsection{Self Reporting}
Collecting longitudinal patient-reported outcomes is known as experience sampling or ecological momentary assessment, a method increasingly implemented via mobile devices \citep{van_berkel_experience_2017}. In general, we refer to this as \emph{self-reporting} or \emph{self-assessment} in this paper.

There are some notable issues with self-reporting. \citet{blome_measuring_2016} outline types of bias in self-reporting: retrospective reports (collected only after an intervention) are vulnerable to recall bias \citep[i.e.\ memory effects;][]{van_den_bergh_accuracy_2016}, while prospective reports (collected before and after an intervention) are susceptible to response shift, in which a respondent changes their interpretation of the scale itself \citep{schwartz_clinical_2006}.
Subjective phenomena such as mood, pain and well-being are typically recorded on ordinal (or Likert) scales, which bring additional analytical challenges, including erroneously treating data as nominal or metric \citep{liddell_analyzing_2018} or floor and ceiling effects \citep{simkovic_robustness_2019, sayers_analysis_2020}, leading to spurious conclusions.

\citet{wang_investigating_2008} investigate the effects of ceiling data in longitudinal analyses. Simulations and empirical data show that ceiling effects can lead to biased parameter estimates and incorrect model selection in growth curve models. The Tobit growth curve model produces more accurate estimates by making the best use of the ceiling data information. This method proves particularly useful in studies where participants frequently or many participants simultaneously hit the upper limit of a scale.


For example, \citet{dixon_how_2019} collected prospective reports of daily pain level. Still, the modeled response in the analysis was a day-to-day change in pain, requiring two consecutive samples to derive a single outcome, a common limitation of measures of variability in prospective self-reports \citep[see, e.g.][]{mun_investigating_2019} that would mean only data from the most engaged participants is preserved \citep{druce_recruitment_2017}.

More data efficient would be for participants to report this change directly, thereby requiring half as many samples and limiting scale attenuation to differences in score rather than their absolute value.
On the other hand, recording only changes in ratings means the baseline magnitude is unknown.
However, inter-individual subjectivity means that such a quantity is not necessarily meaningful between participants and would typically be treated as a random effect in a proper analysis.

\citet{tin_pain_2023} compared different anchors for ordinal pain ratings, with `extremely severe' showing a better correlation with other metrics than boundary anchors such as `as bad as you can imagine.'
Various authors show numerical rating scales (e.g., \ 0--10) are preferred to visual analog scales \citep{firdous_how_2021}.

\citet{matthews_similarities_2018} compared retrospective questionnaires and prospective diary estimates of sleep duration with polysomnography and actigraphy measurements. They found that self-assessed sleep durations were generally 20-30 longer than PSG- and actigraphy measurements.

\citet{tang_JudgementSleepQuality_2023} found people's ratings of their previous night's sleep quality changes throughout the day, influenced by post-sleep factors, including physical activity.

Past work comparing retrospective and prospective self-reports include \citet{fischer_capturing_1999}, which found retrospective and serial measures of change in pain and disability do not give concordant results.
\citet{teirlinck_2019} examined the reliability of retrospective and daily pain measurements.
\citet{oreel_ecological_2020} found ecological momentary assessment captured change in health-related quality of life better than retrospective measures.

However, here we are interested in momentary retrospective assessments; i.e., \ we ask participants every day to compare their experience to yesterday, rather than asking them at the end of a longer time interval to summarize their trajectory over several days, weeks, or months.


\subsection{Working Memory}
Working memory is a core concept in cognition, describing a complex cognitive system with limited capacity for storing and manipulating information that plays an extensive role in human thought process~\cite{Baddeley2003WorkingMemory}.
A review of the literature on the effect of sleep on working memory concludes that working memory performance declines following sleep deprivation~\cite{Frenda2016WMSleep}.
Frenda and Fenn furthermore show that while the consumption of caffeine or other stimulants can be used to temporarily halt performance decrements as the result of sleep deprivation, such approaches are not recommended due to their adverse effects~\cite{Frenda2016WMSleep}.

Given the central role in our cognitive system, working memory inevitably plays a role in our interaction with technology.
HCI research has, for example, investigated the role of notification interruptions on user's working memory~\cite{Oulasvirta2004LTWM} and the impact of working memory on the quality of self-report data~\cite{Berkel2019ContextESM}.
In this work, we are assessing participants' working memory performance in relation to their sleep quality and self-reports.

A widely used task for assessing working memory performance is the `\textit{N}-Back' task, as devised by Kirchner in 1958~\cite{kirchner1958age} and subsequently iterated upon.
During an \textit{N}-Back task, participants are shown a sequence of stimuli and asked to indicate when the current stimulus matches the stimulus shown \textit{n} stimuli ago.
For example, for an \textit{2}-back task and the following set of stimuli characters, the characters in bold indicate a match: Y-O-K-\textbf{O}-H-A-M-\textbf{A}.
A higher \textit{n} corresponds to a higher task difficulty.
Commonly used in Experimental Psychology, \textit{N}-Back tasks have also been employed in HCI research, for example, by Nith et al.\ when investigating the impact of offloading physical tasks on mental workload~\cite{Nith2024SplitBody}.
Critically important to our work, the \textit{N}-back test is a widely used measure for assessing working memory function~\cite{Owen2005Nbackparadigm}.
While there is disagreement whether N-back results yield themselves for inter-individual comparisons \cite{jaeggi_ConcurrentValidityNback_2010, schmitter-edgecombe_CapturingCognitiveCapacity_2024}, in this work we use the scores only for within-subject analysis.

\citet{miyata_PoorSleepQuality_2013} examine the connection of sleep duration and quality with cognitive performance by using the $N$-back test and the Continuous Performance Test (CPT) on 78 adults over 60 years old. They find that the $N$-back test results correlate to sleep duration, efficiency and age. In contrast to our work, they evaluate $N$-back performance only on accuracy, but not on speed, which may be an important factor~\cite{kuriyama_SleepAcceleratesImprovement_2008}. They also do not examine the relationship of the considered measures to self-reports.\citet{vandijk_CognitiveFunctioningSleep_2020} look at the connection between several measures on the Insomnia Severity Index (ISI) and neuropsychological tests, including $N$-back in a working population, but find only that insomnia as assessed by the ISI is significantly related to subjective, but not objective, cognitive performance. In contrast to our work, they look at neither sleep-macro-measurements nor physiological measurements but rely completely on questionnaires.

In this study, we are interested in capturing participants' working memory to indicate their cognitive performance abilities after sleep in an in-the-wild setting. At the same time, the measurement needs to be objective and repeatable every day. The N-back test satisfies the theoretical requirement for measuring working memory while also being conducive to a mobile computing environment. \cite{schmiedek_TaskTaskTask_2014, schmitter-edgecombe_CapturingCognitiveCapacity_2024}

\begin{figure*}[htpb]
  \includegraphics[width=0.33\textwidth]{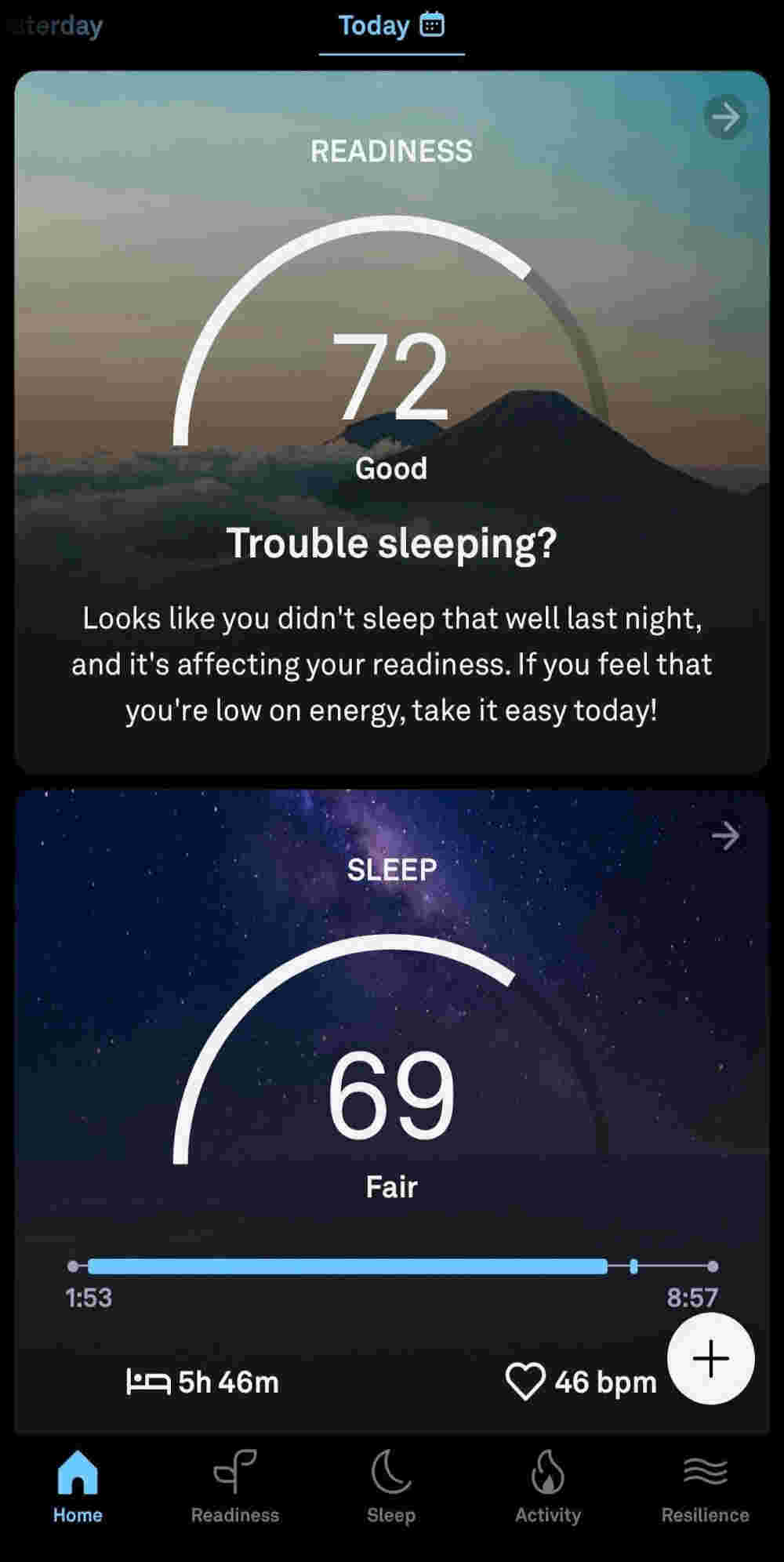}
  \includegraphics[width=0.33\textwidth]{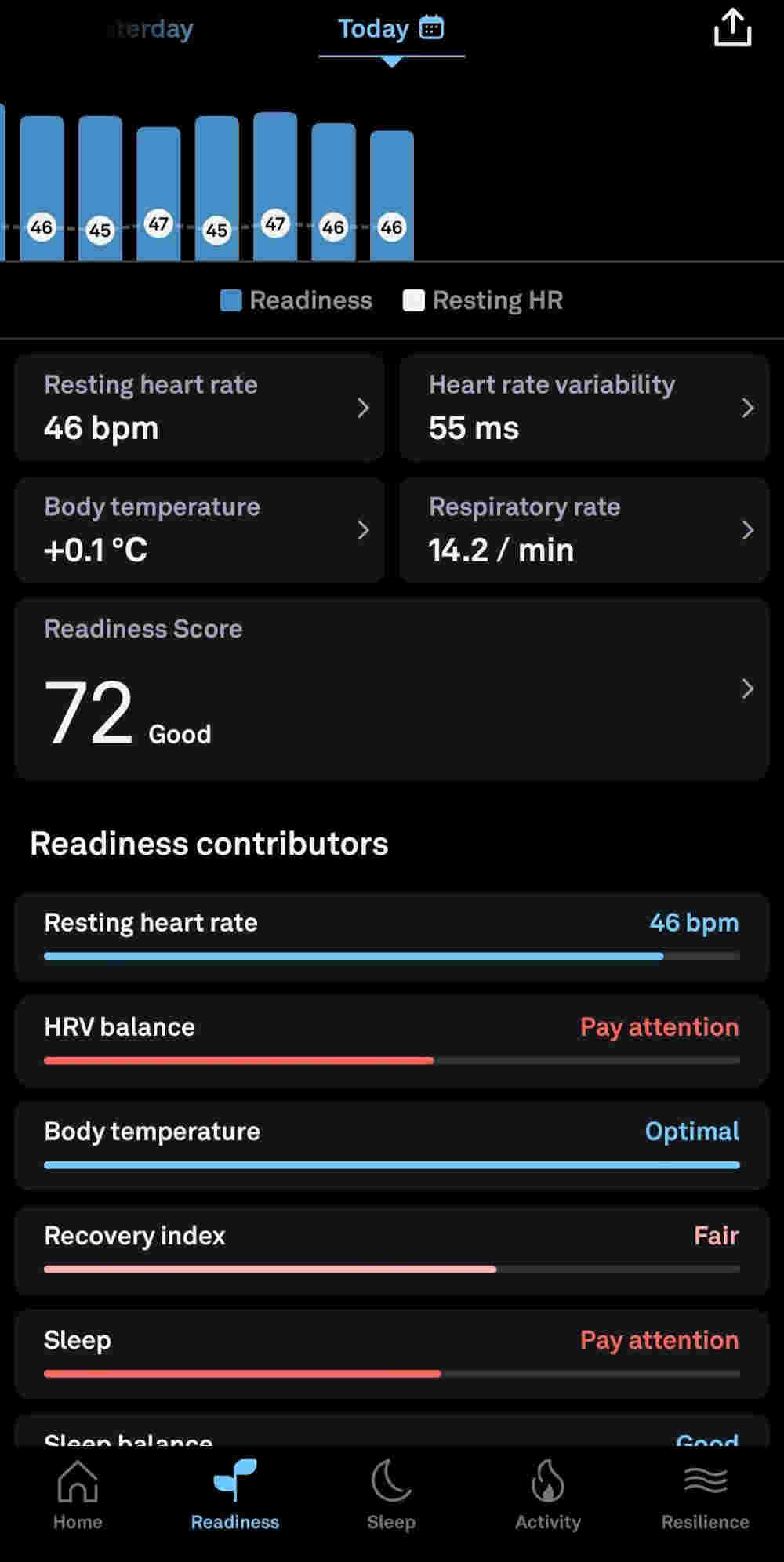}
  \includegraphics[width=0.33\textwidth]{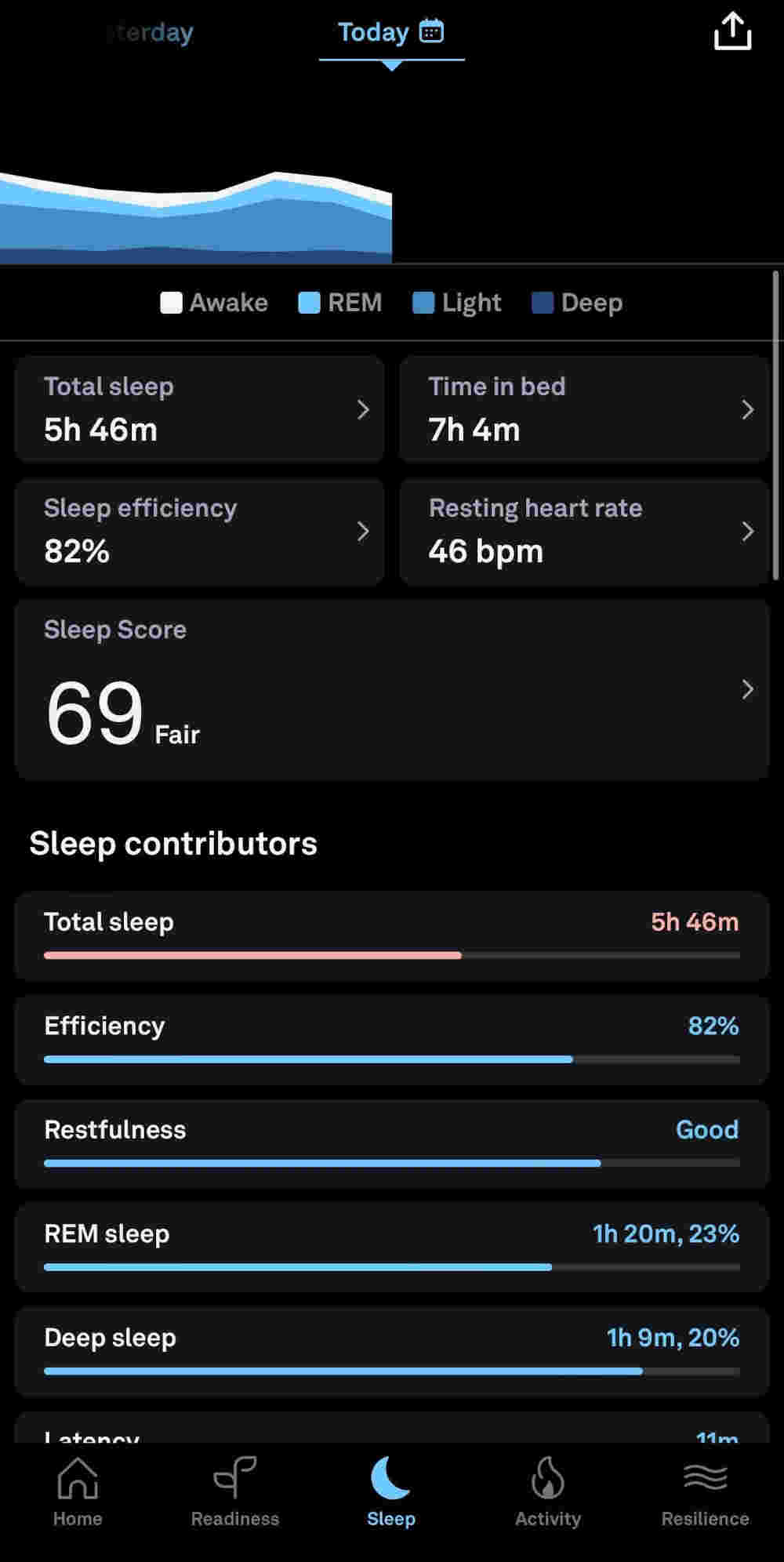}
  \caption{Examples of Data Presented in the Oura Ring Application. On the left is summary data that gives an overview of both sleep quality and readiness. The middle figure provides the data details for readiness, while the figure on the right gives the details for sleep.}
  \Description{The figure contains three images with different screenshots from the Oura app. The purpose of the depiction is to convey how users interact with the sleep data from their sleep tracker (Oura ring).}
  \label{fig:oura-app}
\end{figure*}

\subsection{Alignment of Perceived and Measured Sleep Quality}

\citet{alecci_MismatchMeasuredPerceived_2023} examined the relationship between perceived and measured sleep quality in a cohort of 16 participants over one month, using the M2Sleep dataset by \citet{gashi_RoleModelPersonalization_2022}. Using the Pittsburgh Sleep Quality Index (PSQI), data from Empatica E4 (accelerometer, HR, temperature, and EDA), and MiBands, they found that the MiBand predicts perceived sleep quality worse than a random guesser and concluded that there is a mismatch between measured and perceived sleep quality. They hypothesize this result may change with other devices, such as the Oura ring, which we use in this study. In contrast to their work, by using the Oura ring, we can include sleep phases and measured bedtimes in our analysis.

\citet{cudney_InvestigatingRelationshipObjective_2022} surveyed literature examining the relationship between objective sleep quality measures and self-reported sleep quality. They identified inconsistent results in the surveyed 13 papers: Total sleep time (TLT) and sleep efficiency (SE) were found to be beneficial to sleep quality in some works and detrimental in others, prompting more research, especially in naturalistic settings, which our study offers.

\citet{abdalazim_BiHeartSBilateralHeart_2023} released a dataset of 10 participants over 30 consecutive days, where participants gave sleep-quality self-reports in the morning and evening. During the night, participants wore Oura rings on their left hand and Empatica E4 wristbands on both hands, collecting physiological data like HR, HRV, bed and sleep times, inter-beat interval (IBI), blood volume pulse (BVP), electrodermal activity (EDA) among others. The main differences in our work are that we included and analyzed working memory scores obtained from the N-back test, as well as a larger participant number and experiment duration period. Additionally, the self-reports in our work are carried out in a comparative way to the previous night.

To our knowledge, this is the first work combining perceived sleep quality, objective sleep measures (including sleep phases) and working memory scores.

\section{Experiment}

To test the hypothesis, we developed an experiment smartphone application and recruited participants to complete the test suite daily. The app consists of a sleep self-reporting question and an \textit{N}-back ($N=3$) test of length 20 as an indicator for working memory capacity. We also record participants' sleep data. The participants are asked to do this while following their otherwise regular schedules to obtain readings corresponding to their usual behaviors and environments. They are, therefore, closer to a real-world application. The experimental procedures for this study were reviewed and approved by the university faculty’s Ethics Committee. Informed written consent was obtained from all participants before their involvement in the study.

\subsection{Sleep Data}\label{sec:sleepdata}

We use the Oura ring, a sleep and fitness tracker device with a ring form factor, to record sleep data. We decided on the Oura Ring due to its form factor, battery life, and integration of user-facing software.

The Oura ring continuously gathers physiological data using red, green, and infrared LEDs alongside an NTC temperature sensor and a 3D accelerometer. The LEDs monitor blood vessel dilation, contraction, and blood oxygen levels (SpO2) using photoplethysmography, with a sampling rate of 250 times per second. According to Oura, this achieves ``99.9\% reliability compared to a medical-grade electrocardiogram''~\cite{_TechnologyOuraRing_2020}. These readings derive metrics such as heart rate (HR), heart rate variability (HRV), and breathing rate. While generation 2 rings only monitor HR and HRV during sleep, the generation 3 rings extend this to daytime monitoring. Since only measurements during sleep are considered in our work, generation 2 and 3 rings are used. 
The NTC temperature sensor records skin temperature every minute, but temperature data is the deviation from the average overnight temperature compared to a long-term baseline~\cite{_BodyTemperatureAccessed_}. The 3D accelerometer tracks movement to classify sleep, sleep phases, and restless periods.
The sleep data is cached on the ring and periodically sent to the smartphone. The Oura application then sends the data to Oura's servers, where we can download the participant data.

The sleep features we use are total sleep duration, REM sleep duration, deep sleep duration, light sleep duration, overnight average heart rate (HR), overnight average heart rate variability (HRV), average breathing rate (BR), body temperature deviation from the average, efficiency (total sleep duration as a percentage of time in bed), latency, restless periods, bedtime start and bedtime end. We additionally use the sleep score, an overall sleep measure used by Oura calculated from other features, but the exact relationship is proprietary and unknown. We chose these features because they are the main sleep-related features provided by the Oura ring.

\subsection{Sleep Quality Assessment}

Participants were asked to download a custom smartphone application developed by the authors, available for both Android and iOS. They were instructed to open the app daily within 30 minutes of waking up, but before viewing their sleep data from the Oura app, eating breakfast, consuming alcohol, or engaging in physical activity. No restrictions were placed on their sleep habits, such as when to sleep or wake up.

Upon opening the app, users first select their preferred language, English or Japanese. On the following screen, they must confirm that they have not yet checked their sleep data from the Oura app. This step ensures that participants’ self-assessments are not influenced by the objective data provided by the Oura ring. Once confirmed, participants are asked, “How was your sleep compared to yesterday?” and can respond on a five-point scale: 1) Much better, 2) Better, 3) No Change, 4) Worse, 5) Much worse. (Japanese: \begin{CJK}{UTF8}{min}``昨日と比べて、今日の睡眠の質はどうでしたか？'' with answers 1) ``大幅に良くなっている'', 2) ``やや良くなっている'', 3) ``変わらない'', 4) ``やや悪くなっている'', 5) ``大幅に悪くなっている''\end{CJK}). We chose this type of self-reporting of the perceived sleep quality relative to the previous night to circumvent floor and ceiling effects. The question about sleep quality is purposely vague to capture the full range of what participants perceive as ``sleep quality'' and to make it similar to clinical studies inquiring about, e.g., pain levels, where it is similarly unclear which aspect of the examined property participants focus on. The screen for the self-assessment can be seen in Figure~\ref{fig:experiment-app} on the left.

A one-item sleep quality scale allows us to examine subjective self-assessments of sleep quality and how these align with objective sleep metrics and cognitive performance concisely. Unlike more structured instruments like the PSQI, the single-item approach avoids constraining participants to predefined dimensions of sleep quality (e.g., latency, disturbances) that might not align with their individual experiences. A single item minimizes respondent burden or fatigue, especially in the intensive longitudinal design \citep{song_examining_2023}, and captures participants' holistic judgment of sleep quality.

\subsection{\textit{N}-Back Test}

\begin{figure*}[htpb]
  \includegraphics[width=0.33\textwidth]{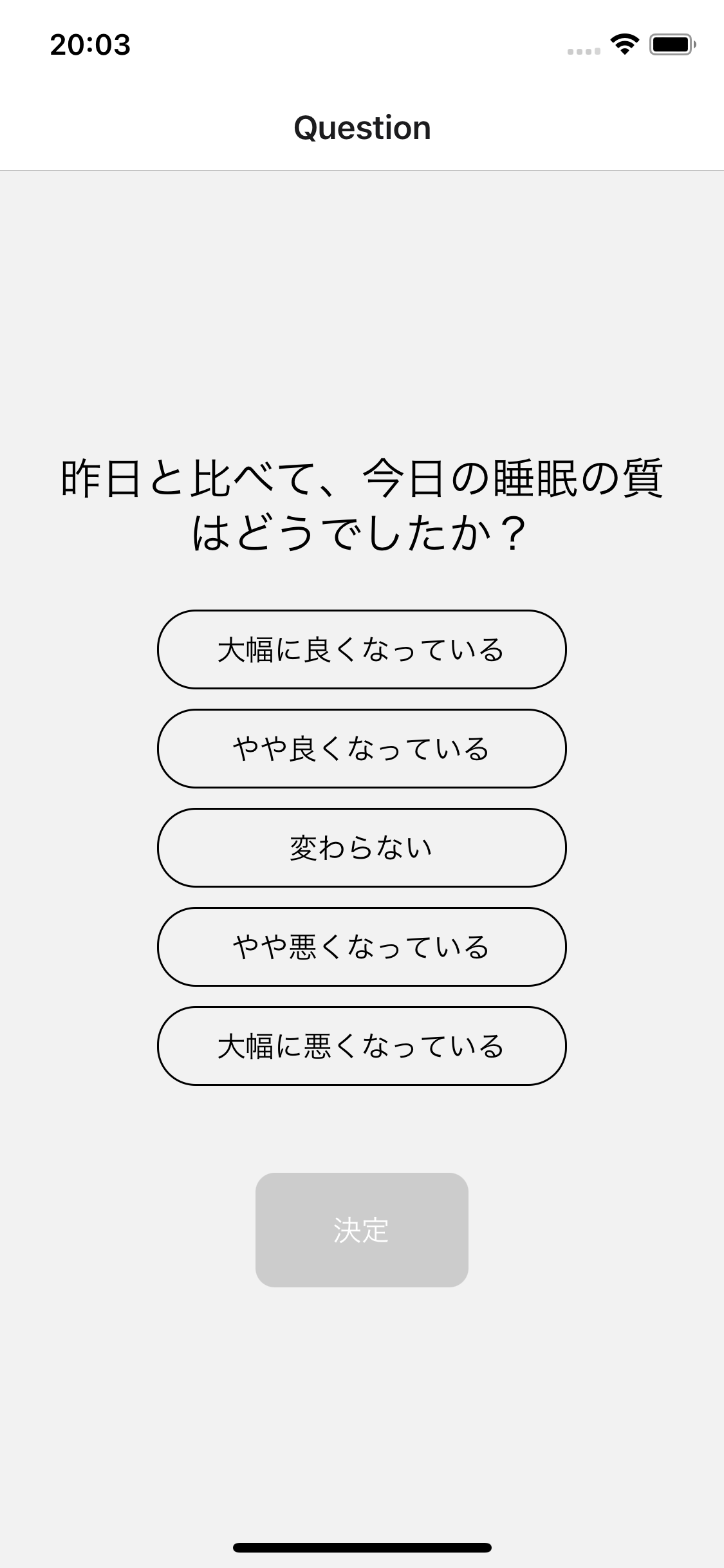}
  \includegraphics[width=0.33\textwidth]{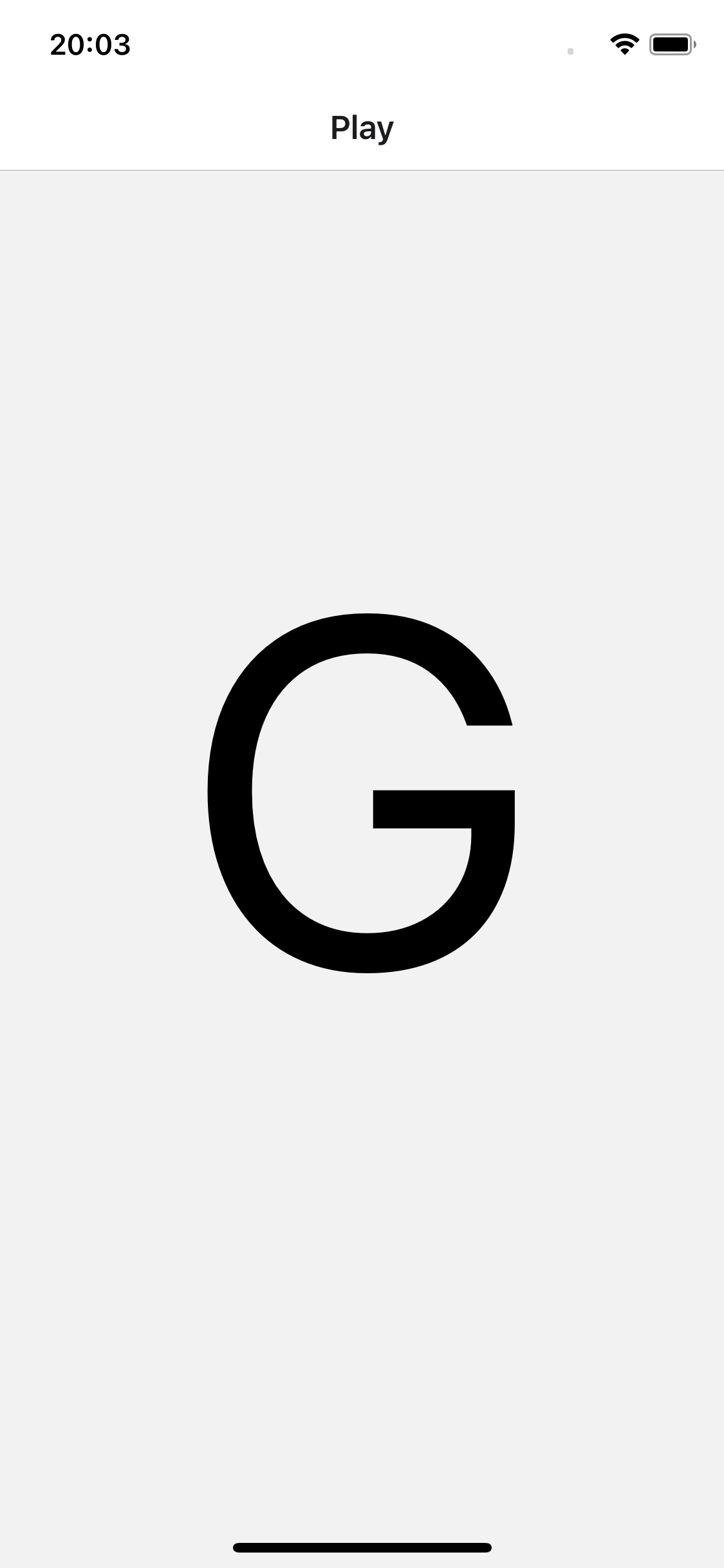}
  \includegraphics[width=0.33\textwidth]{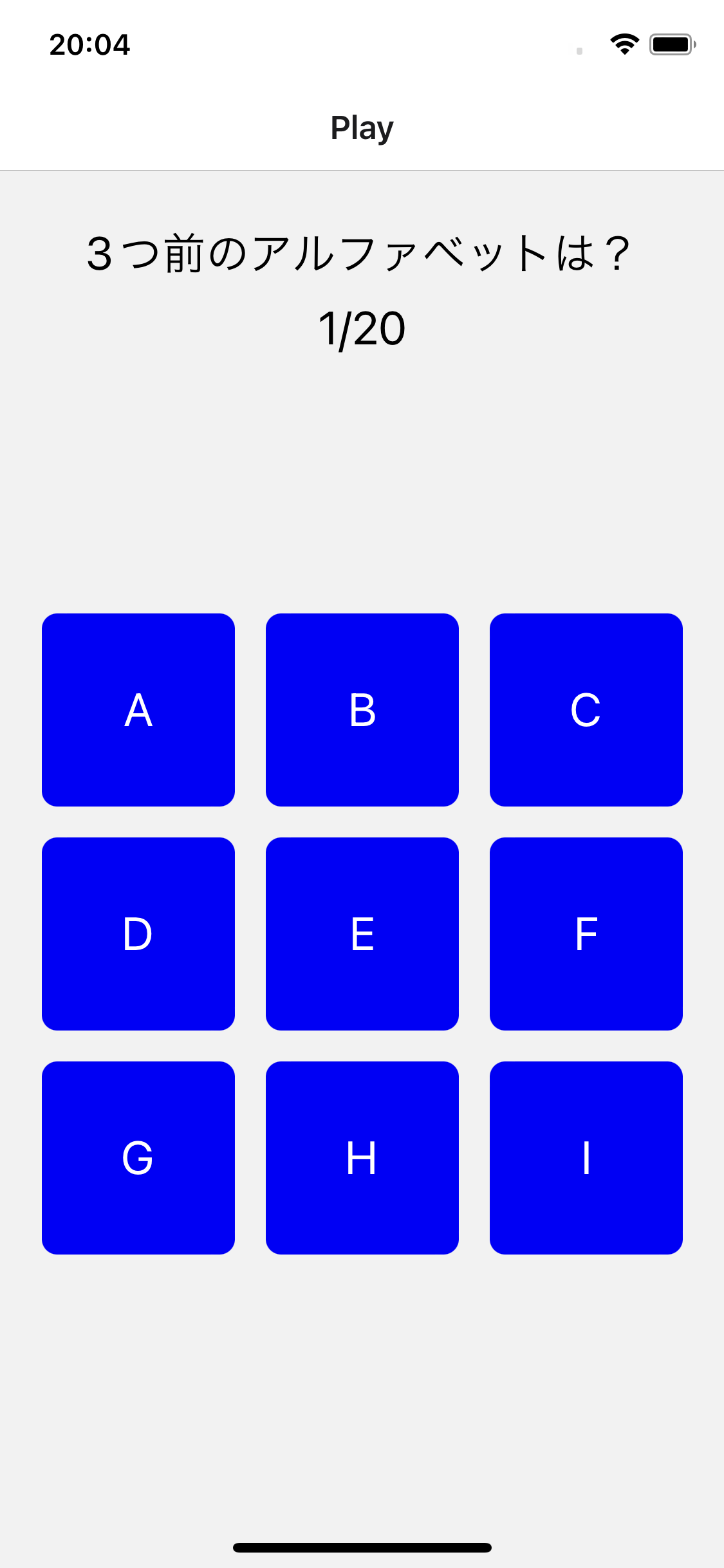}
  \caption{Experiment app. Left: Sleep assessment question. Center: \textit{N}-back showing the letter. Right: User being prompted for input during \textit{N}-back.}
  \Description{The figure shows three screenshots from our experiment app, where every screenshot depicts a different state during the app. The first shows the sleep question, the second a letter showed during the sequence of the \textit{N}-back test, and the third a prompt during the \textit{N}-back test.}
  \label{fig:experiment-app}
\end{figure*}

After responding to the sleep question, participants proceed to an \textit{N}-back test with N=3. In this test, users are shown a sequence of letters from A to H, each displayed on the screen for 1.5 seconds. From the fourth letter, participants are prompted to identify the letter shown three positions earlier by selecting from 9 buttons labeled A through H. This prompt stage has no time limit, but the response time for each prompt is recorded. The sequence continues until the participant has been prompted 20 times, resulting in 23 letters displayed in total. We record both the number of correctly recalled letters and the response times. Participants do not receive feedback on their performance during the experiment. Screenshots of the \textit{N}-back section of the experiment app are shown in Figure~\ref{fig:experiment-app} (center and right).

We adopted the same settings as presented in related studies on sleep and working memory~\cite{kuriyama_SleepAcceleratesImprovement_2008}, i.e., we presented stimuli for 1.5 seconds and required participants to answer 20 questions. 
To determine the appropriate difficulty level for the experiment, measurements were conducted using two types of N-back tasks with varying difficulty levels (N=2, N=3) over three days each, involving two participants. Participants first completed the N=2 N-back task for three days, followed by a five-day interval before undertaking the N=3 N-back task for another three days. The results showed that for the N=2 task, half of the results (not participants) were perfect scores, whereas the N=3 task exhibited some variability in outcomes.
Considering the requirements for long-term measurement, it was essential to select a task that maintains a balance between participant adherence and ensuring variability in results. While more challenging \textit{N}-back tests provide a better measure of working memory capacity, they also reduce the likelihood of participants completing the experiment. Based on this balance, the N=3 task was identified as the most appropriate for the study and was adopted accordingly.

To evaluate accuracy and speed, we calculate a composite score based on the number of correct responses and response times:

\begin{align}
    S &= C + \frac{T_\mathrm{max}-T}{T_\mathrm{max}-T_\mathrm{min}} \;\;,
\end{align}

where $C$ is the percentage of correctly memorized letters, $T$ is the average response time and  $T_\mathrm{max}$ and $T_\mathrm{min}$ are the maximum and minimum response time respectively. Both correctness and time are evaluated in the range [0,1], resulting in $S$ ranging in [0,2].

Performance in the N-back task decreases with age~\cite{gajewski_WhatDoesNBack_2018}; thus, we can expect relatively good scores from our younger participant cohort. By using participant ID as a random effect, we model day-to-day differences in N-back performance relative to the individual and do not compare scores between participants.

\subsection{Participants}
Experiment participants were recruited from two groups of university students. First, 6 participants were recruited from a parallel-running long-term study on wearables at the host university. In this case, participants have volunteered to wear an Oura Ring as part of an in-the-wild study. Use of the wearable, in this case, is not restricted or enforced, meaning that the participant can wear the ring as much or as little as they want. 23 additional participants were recruited from outside the study. These participants were given Oura Rings before the study began and were required to wear the rings for two weeks before the study began to calibrate the Oura Ring's physiological measurements and readiness scores.


Recruitment began in June 2024 on a rolling basis, with the last participant starting on July 17, 2024. Data collection concluded on September 3, 2024. Participant recruitment was open to students of the university who were willing to wear the ring, answer the survey, and complete the N-back tests. Of the 35 participants who started, six dropped out for different reasons (e.g., lost ring, lack of time, or a broken ring). Table~\ref{tab:participants} lists participants’ start and end dates, completed days, consecutive days, and missed days.

\begin{table*}[htpb]
\centering
\caption{Participant start and end date and number of completed days and consecutive day-pairs.}
\label{tab:participants}
\begin{tabular}{rrrrrr}
\toprule
Participant No. & Start Date & End Date & Completed Days & Consecutive Day-Pairs & Days Missed\\
\midrule
1 & 2024-06-14 & 2024-08-31 & 66 & 59 & 12 \\
2 & 2024-06-19 & 2024-08-21 & 59 & 54 & 4 \\
3 & 2024-06-22 & 2024-09-03 & 62 & 53 & 11 \\
4 & 2024-06-14 & 2024-09-01 & 61 & 47 & 18 \\
5 & 2024-06-21 & 2024-08-29 & 57 & 45 & 12 \\
6 & 2024-06-22 & 2024-08-06 & 45 & 44 & 0 \\
7 & 2024-06-19 & 2024-08-05 & 46 & 44 & 1 \\
8 & 2024-07-05 & 2024-09-03 & 52 & 42 & 8 \\
9 & 2024-06-19 & 2024-08-01 & 43 & 42 & 0 \\
10 & 2024-06-26 & 2024-08-29 & 51 & 42 & 13 \\
11 & 2024-06-20 & 2024-08-15 & 47 & 42 & 9 \\
12 & 2024-06-21 & 2024-08-14 & 48 & 41 & 6 \\
13 & 2024-06-10 & 2024-09-03 & 49 & 41 & 36 \\
14 & 2024-06-18 & 2024-09-01 & 57 & 41 & 18 \\
15 & 2024-06-22 & 2024-08-20 & 49 & 40 & 10 \\
16 & 2024-07-17 & 2024-08-31 & 43 & 40 & 2 \\
17 & 2024-07-10 & 2024-09-03 & 47 & 40 & 8 \\
18 & 2024-06-23 & 2024-08-31 & 49 & 39 & 20 \\
19 & 2024-06-14 & 2024-08-11 & 47 & 38 & 11 \\
20 & 2024-06-27 & 2024-09-03 & 48 & 38 & 20 \\
21 & 2024-06-18 & 2024-09-03 & 49 & 37 & 28 \\
22 & 2024-06-21 & 2024-08-10 & 43 & 37 & 7 \\
23 & 2024-06-25 & 2024-08-16 & 43 & 36 & 9 \\
24 & 2024-07-10 & 2024-09-03 & 42 & 31 & 13 \\
25 & 2024-06-19 & 2024-09-03 & 39 & 31 & 37 \\
26 & 2024-07-07 & 2024-08-19 & 35 & 28 & 8 \\
27 & 2024-06-23 & 2024-09-03 & 43 & 28 & 29 \\
28 & 2024-07-13 & 2024-09-03 & 39 & 26 & 13 \\
29 & 2024-07-10 & 2024-08-22 & 33 & 25 & 10 \\
\bottomrule
\end{tabular}
\end{table*}

\subsection{Experiment Procedure}

Each participant attended a brief introductory session where the experiment app, sleep questionnaire, and \textit{N}-back test were explained, along with the experiment rules. New Oura ring users were assisted with ring sizing~\cite{neigel_ExploringUsersAbility_2023}, Oura ring setup, app installation, and subscription activation. Participants completed a test run of the entire experiment app; these results were discarded. Participants were instructed to start the experiment the following day.

Participants could choose to participate for 4, 6, or 8 weeks. A completed week was defined as seven consecutive days of complete data, including data for the preceding day. For example, full data provided on June 20, 21, 23, and 24 would count as four completed days and two consecutive days. Participants were remunerated 1250 yen for each set of 7 consecutive completed days. If participants did not meet their agreed quota by the final data collection day (September 3, 2024), they were paid pro rata based on the number of consecutive days they completed.

A day was considered missed if the experiment app suite was not completed or the Oura ring was not worn overnight. Any day with complete data, where the previous day’s data was also complete, counted as a consecutive day.

After the experiment, participants filled out a short post-experiment survey, where they answered two questions and could give additional comments about the experiment. The questions were 1.) ``Do you think the Oura ring mostly confirmed or contradicted your own feeling about your sleep?'' (\begin{CJK}{UTF8}{min}リングが提供する睡眠データは、あなたの感覚と一致していましたか？それとも一致していませんでしたか？\end{CJK}) and 2.) ``How long did it take for you to get familiar with the \textit{N}-back test?'' (\begin{CJK}{UTF8}{min}\textit{N}-Backテストに慣れるまでどのくらい時間がかかりましたか？\end{CJK}).

For every participant, we exclude the first few days of \textit{N}-Back data, according to the days they reported they needed to get used to the test. Sleep self-assessment is not influenced by this. Values for the familiarization with the \textit{N}-back test ranged from 0 days (immediately after the first test run) to 10 days, with an average of 4.3 days (2.8 STD). One participant didn't take the exit survey, so 4 days of familiarization were assumed for them. One participant misunderstood the \textit{N}-Back experiment to be of $N=2$. This was noticed after 14 days, and after another explanation, the participant continued. The first 14 days \textit{N}-Back results are not considered for this participant. Their sleep self-assessment is not influenced by this.

\begin{figure*}[htpb]
\centering
  \noindent\includegraphics[width=0.95\textwidth]{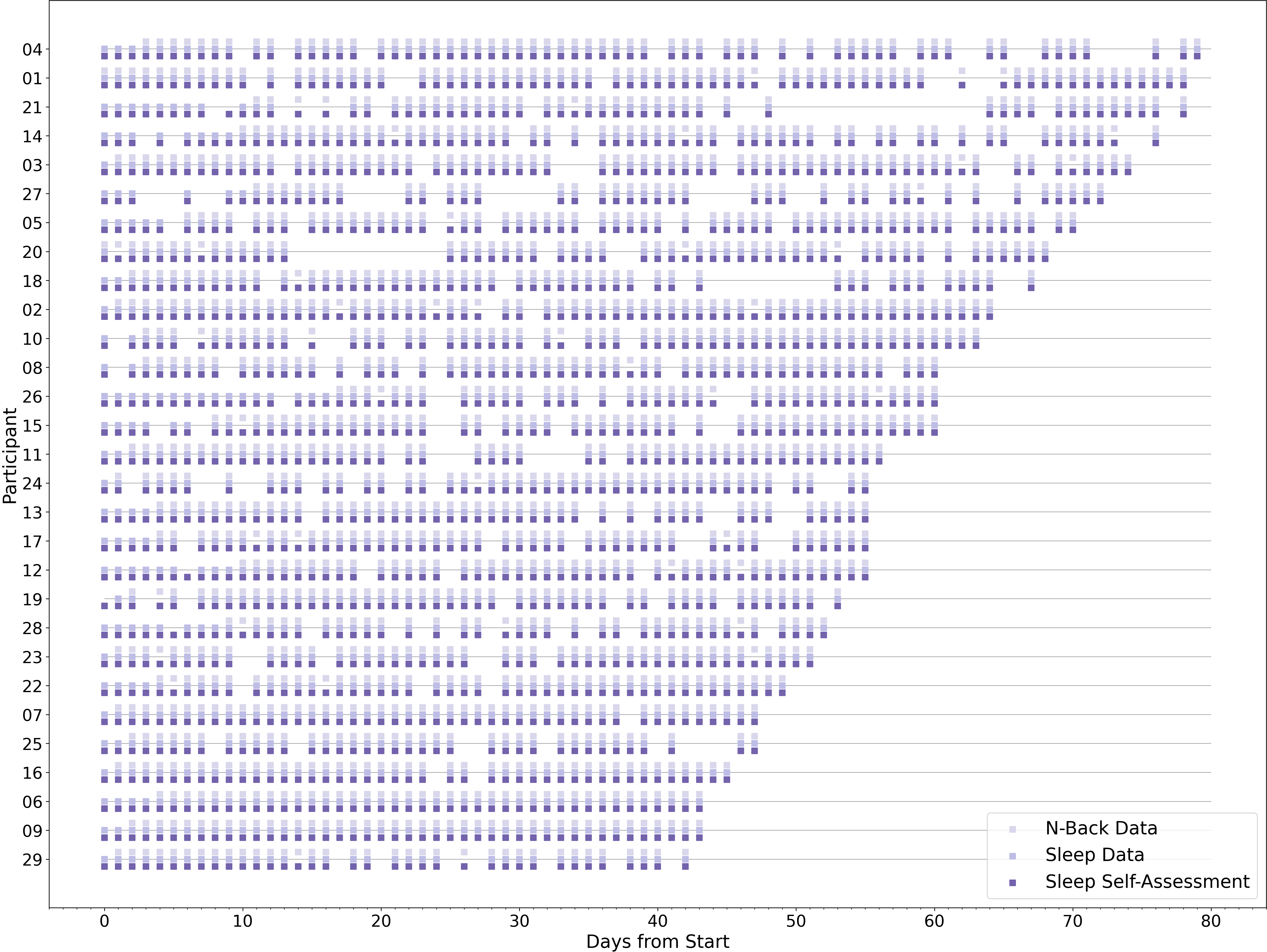}
  \caption{Data availability per participant against days from experiment start.}
  \Description{The image shows which data type is available for which day and for which participant. The data types are sleep self-report, \textit{N}-back score, and sleep tracker data. It is visible that experiment durations are not equal among participants. It's also visible that participants have gaps in their data reporting, i.e. days where they didn't do the experiment or didn't wear the sleep tracker}
  \label{fig:data_availability}
\end{figure*}

\begin{figure*}[htpb]
\centering
  \includegraphics[height=0.9\textheight]{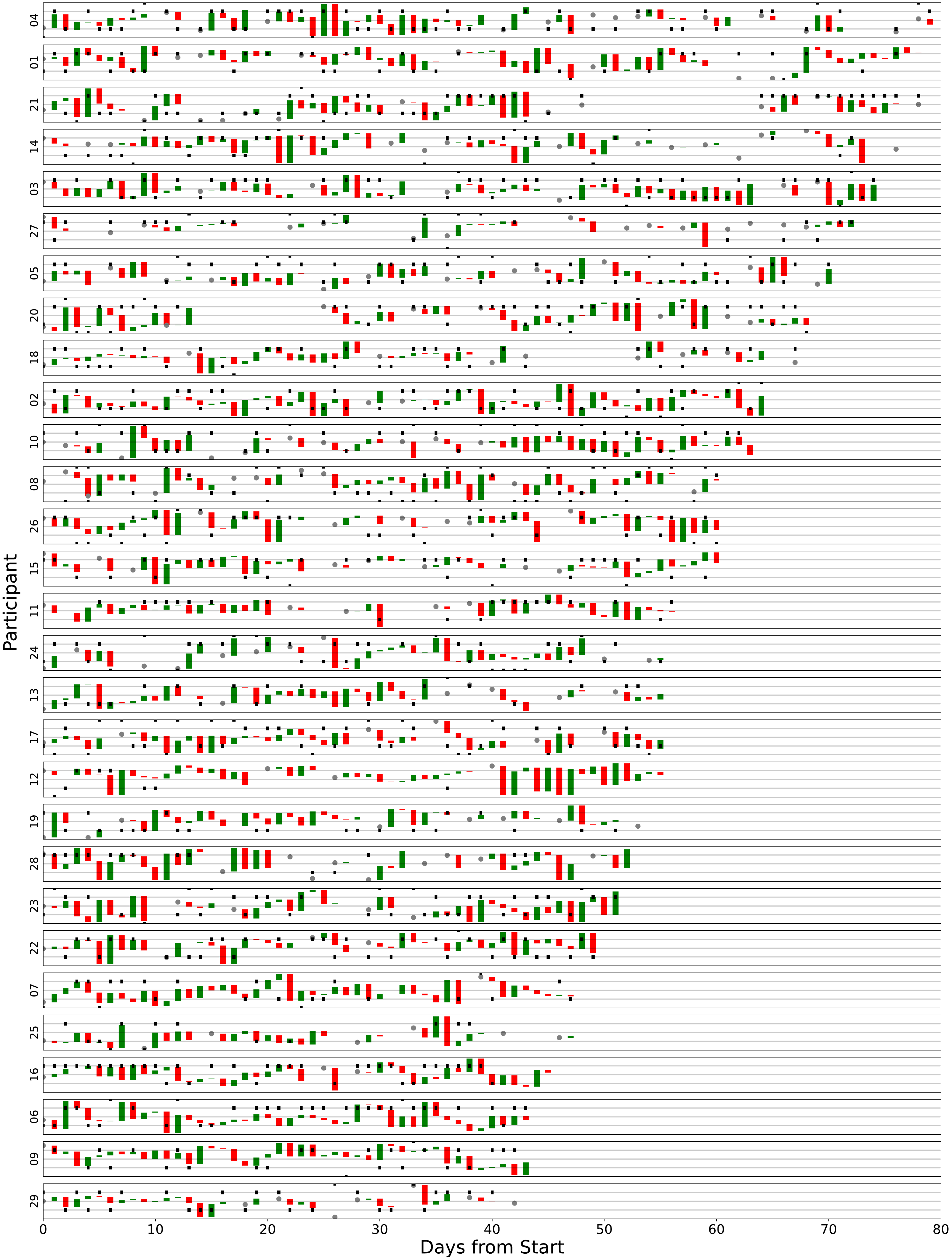}
  \caption{Change in REM sleep duration for participants and their self-assessment. Green and red candlesticks depict the increase and decrease in the absolute level of REM sleep, while grey dots show only the absolute level if no data is available for the previous day. Black squares arranged on five horizontal lines indicate the sleep self-assessment for that day.}
  \Description{The shape of the figure resembles that of Figure 3 (the data availability figure), but here, for each day and participant, it illustrates 1) changes in their REM sleep duration and 2) their sleep self-assessment.}
  \label{fig:candlesticks}
\end{figure*}

Of the 29 participants, 22 identified as male and seven as female. The participants' mean age was 22.6 (2.7 STD). Data availability (self-assessment, sleep data, and \textit{N}-Back data) per participant can be seen in Figure~\ref{fig:data_availability}. Changes in total sleep duration and self-assessment can be seen in Figure~\ref{fig:candlesticks}.

\section{Analysis \& Results}

In total, the dataset contains 1413 days with valid \textit{N}-Back results, 1546 days with self-reported sleep quality, 1679 days with sleep data, 1367 with sleep data, and valid \textit{N}-Back results (and therefore complete data) and 1478 with self-reported sleep quality and sleep data. Further, there are 1158 consecutive day pairs with valid \textit{N}-back results and sleep data, 1257 days with consecutive self-reported sleep quality and sleep data, and 1158 consecutive day pairs with complete data available. On average, participants took $58 \pm 34$ seconds in total for completing the \textit{N}-Back prompts (excluding the 1.5 seconds that the letter is shown). Out of 29 participants, 28 filled out the exit survey. Out of these, 25 reported that they felt their sleep data mostly confirmed their feeling about their sleep, while the other three reported that the ring mostly contradicted their own feeling about their sleep.

\subsection{General Correlations between Sleep Markers, Sleep Self-Reports and \textit{N}-Back Scores}\label{sec:corr}

In addition to the sleep features described in Section~\ref{sec:sleepdata}, we calculate sleep phase percentages as the duration of each sleep phase divided by the total sleep duration. Bedtime start is represented numerically as the floating-point value of hours and minutes past 17:00, while bedtime end is represented as hours and minutes past midnight. All sleep durations are expressed in hours, and sleep latency is in minutes. Restless periods are expressed as a percentage of total sleep duration, as their duration was highly correlated with total sleep duration.

To investigate how sleep self-reports align with changes in sleep markers, we first examine Spearman’s rank correlations between sleep features, \textit{N}-back results, and self-reports. We use Spearman's rank correlation here because self-reports are on a Likert scale and do not represent a real numerical value. Since self-reports reflect changes compared to the previous night, we used the difference in each feature from the previous night’s value. Only days with consecutive data for the respective feature were considered. The results are shown on the left side of Table~\ref{tab:combined-correlation}.

We found significant but weak correlations between self-assessment and the difference in Oura sleep score ($0.38$), total sleep duration ($0.34$), and REM sleep duration ($0.32$). There were also significant but very weak correlations with changes in light sleep duration ($0.29$), deep sleep duration ($0.22$), REM percentage ($0.23$), sleep efficiency ($0.15$), and bedtime end ($0.21$). Additionally, we observed significant but weak inverse correlations with changes in deep sleep percentage ($-0.14$) and bedtime start ($-0.21$). These findings suggest that self-assessment improves with increased sleep duration (and consequently, later bedtime) and efficiency. The inverse correlation with deep sleep percentage is likely because sleep phase percentages are inversely related: if REM percentage increases, deep and/or light sleep percentages must decrease. This, along with REM sleep having the highest coefficient among sleep phases, suggests that REM sleep is most closely linked to self-assessment.

We find a statistically significant but negligible correlation between self-reports and difference in \textit{N}-Back score ($0.08$).

To verify whether \textit{N}-Back results as a measure of cognitive performance are related to sleep measures, we check Spearman's Rank correlations of sleep features with the \textit{N}-Back score. We use Spearman's Rank correlation here because most features exhibit skewness and outliers, confirmed by low Shapiro-Wilk test statistics. The results are displayed in Table~\ref{tab:combined-correlation} on the right.

\begin{table*}[htpb]
\centering
\caption{Spearman's rank correlation of self-assessment (left) and Pearson correlation of \textit{N}-Back score (right) to sleep features. Stars denote $p$-values smaller than $0.05, 0.01, 0.001$ with Bonferroni correction.}
\label{tab:combined-correlation}
\begin{tabular}{l r r l r r}
\toprule
\multicolumn{3}{c}{Self-Assessment} & \multicolumn{3}{c}{\textit{N}-back Score} \\
\cmidrule(r){1-3} \cmidrule(l){4-6}
Feature Difference & {Spearman's Rank} & {$p$-value} & Feature & {Spearman's Rank} & {$p$-value} \\
\cmidrule(r){1-3} \cmidrule(l){4-6}
Sleep Score & 0.380 & <0.001*** & Sleep Score & 0.004 & 0.878 \\
Total Sleep & 0.340 & <0.001*** & Total Sleep & 0.075 & 0.004 \\
REM Sleep & 0.320 & <0.001*** & REM Sleep & 0.036 & 0.162 \\
Deep Sleep & 0.220 & <0.001*** & Deep Sleep & 0.050 & 0.053 \\
Light Sleep & 0.290 & <0.001*** & Light Sleep & 0.069 & 0.008 \\
Avg. HR & $ -0.070$ & 0.009 & Avg. HR & $-0.016$ & 0.557 \\
Avg. HRV & 0.040 & 0.104 & Avg. HRV & $-0.050$ & 0.058 \\
Avg. BR & $ -0.060$ & 0.023 & Avg. BR & 0.066 & 0.012 \\
Temp. Deviation & 0.020 & 0.487 & Temp. Deviation & 0.006 & 0.839 \\
REM Percentage & 0.230 & <0.001*** & REM Percentage & 0.004 & 0.865 \\
Deep Percentage & $ -0.140$ & <0.001*** & Deep Percentage & $-0.020$ & 0.451 \\
Light Percentage & $ -0.030$ & 0.256 & Light Percentage & 0.007 & 0.787 \\
Efficiency & 0.150 & <0.001*** & Efficiency & $-0.071$ & 0.007 \\
Latency & $ -0.020$ & 0.413 & Latency & 0.024 & 0.360 \\
Restless Periods & 0.030 & 0.263 & Restless Periods & 0.132 & <0.001*** \\
Bedtime Start & $ -0.210$ & <0.001*** & Bedtime Start & 0.059 & 0.025 \\
Bedtime End & 0.210 & <0.001*** & Bedtime End & 0.116 & <0.001*** \\
\textit{N}-Back Score & 0.080 & 0.003* & Self-Assessment & 0.060 & 0.018 \\
\bottomrule
\end{tabular}
\end{table*}

We found significant but negligible correlations between \textit{N}-back scores and the number of restless periods (0.132) as well as bedtime end (0.116). Notably, no correlations were found between sleep (phase) durations and \textit{N}-back performance.

\subsection{Mixed Effects and Ordinal Models}

To evaluate whether participants’ self-assessments have predictive value for combinations of sleep markers, we applied reverse regression~\cite{zhang_reverse_2014}, modeling self-assessment using sleep markers and \textit{N}-back scores.

The self-assessment responses, being qualitative and aligned on a Likert scale, are ordinal. However, since they theoretically represent a first-order derivative of subjective sleep quality, they can be mapped symmetrically around zero, e.g., $+2x$ for “Much better,” $+x$ for “Better,” $0$ for “No change,” $-x$ for “Worse,” and $-2x$ for “Much worse,” where $x$ is arbitrary.

Given that some features are highly correlated or derived from others, we used a reduced feature set to model self-assessment with a linear mixed effects model, incorporating participant ID as a random effect to account for inter-individual variability. Starting from the full feature set, features were iteratively removed to minimize variance inflation factor (VIF)~\cite{thompson_ExtractingVarianceInflation_2017}  and Akaike information criterion (AIC)~\cite{akaike_AkaikesInformationCriterion_2011}. Once all VIF values were below 5, further reductions were made by removing features with high $p$-values if doing so lowered AIC. The results of the optimized model are shown in Table~\ref{tab:LME-selfassessment}. The key features with the highest predictive quality are differences in average HR, REM sleep duration, bedtime start and end, and \textit{N}-back scores. This model explains 15\% of the variance in sleep self-reports through fixed effects and 18\% when including random effects.

\begin{figure*}[htpb]
\centering
  \includegraphics[width=0.8\textwidth]{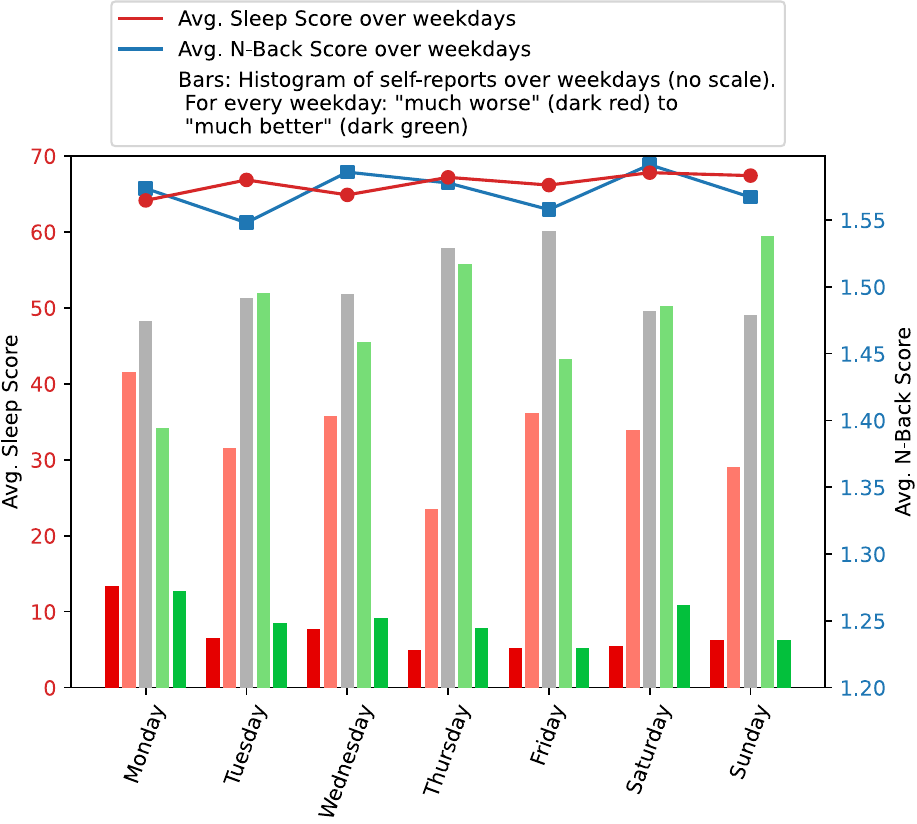}
  \caption{Histograms for all five self-report answer options displaying their relative frequency against the days of the week, overlaid with two line plots showing the average sleep score (red line with circular markers, left axis) and average N-Back score (blue line with square markers, right axis). Bars of any one color, e.g., deep red, sum to one.}
  \Description{The figure has weekdays Monday through Friday on the x-axis and depicts three things: 1) Histograms for sleep self-assessment. The histograms show the relative frequency of every possible answer over the weekdays, i.e., there are 7 bars for the option ``much worse'', one for each weekday respectively, and they all add to 1. The same is true for the other self-assessment options. It can be seen that sleep self-assessment tends to be worse on Mondays and better on Thursdays and Sundays. 2) A line plot showing the average sleep score for the cohort over the weekdays. It can be seen that the sleep score is lower on Mondays. 3) A line plot showing the average \textit{N}-back score over the weekdays. It can be seen that Tuesdays, Fridays, and Sundays have lower scores. }
  \label{fig:selfa_weekday}
\end{figure*}

Since sleep scores tend to be slightly lower on Mondays and higher on weekends, see Figure~\ref{fig:selfa_weekday}, we try to include the day of the week as a random effect in our mixed effects model. We find the covariance of the day of the week to be of order $10^{-9}$ and the identified coefficients of the sleep feature to be equal to a model without the day of the week as a random effect, so we remove it.
The uneven data availability per participant is mostly handled by the mixed effects model's random effect. To ensure it has no significant effect, we ran the mixed effects model again with participants 1. (most data) and 29. (least data) removed. We find that compared to the full participant base model, coefficients differentiate between 0\% and 20\%, with avg. HR difference and bedtime end hour difference unchanged, and bedtime start hour difference showing the biggest difference (-0.04 for full participant base vs. -0.048), see Table~\ref{tab:LME-selfassessment-reducedpart}. This confirms that the uneven data distribution does not significantly alter the results.

\begin{table}[htpb]
\centering
\caption{Mixed Effects Model for Self-Assessment.}
\label{tab:LME-selfassessment}
\resizebox{\columnwidth}{!}{%
\begin{tabular}{lrrr}
\toprule
Variable & Coefficient  & Std. Err. & $p$-value \\
\midrule
Intercept & 0.120 & 0.040 & 0.003** \\
Avg. HR Difference & $ -0.016$ & 0.005 & 0.001** \\
REM Sleep Duration Difference & 0.281 & 0.046 & <0.001*** \\
Bedtime End Hour Difference & 0.050 & 0.013 & <0.001*** \\
Bedtime Start Hour Difference & $ -0.040$ & 0.013 & 0.001** \\
\textit{N}-Back Score Difference & 0.518 & 0.144 & <0.001*** \\
\multicolumn{4}{r}{\footnotesize * $p<0.05$, ** $p<0.01$, *** $p<0.001$}\\
\multicolumn{4}{r}{\footnotesize $R^2_{\text{marginal}}=0.15$, $R^2_{\text{conditional}}=0.18$}\\
\bottomrule
\end{tabular}}
\end{table}

\begin{table}[htpb]
\centering
\caption{Mixed Effects Model for \textit{N}-Back Score.}
\label{tab:LME-nback}
\resizebox{\columnwidth}{!}{%
\begin{tabular}{lrrr}
\toprule
Variable & Coefficient  & Std. Err. & $p$-value \\
\midrule
Intercept & 1.530 & 0.034 & <0.001*** \\
REM Sleep Duration & 0.033 & 0.012 & 0.007** \\
REM Sleep Duration Difference & $ -0.022$ & 0.008 & 0.004** \\
Self-Assessment & 0.017 & 0.005 & <0.001*** \\
\multicolumn{4}{r}{\footnotesize* $p<0.05$, ** $p<0.01$, *** $p<0.001$}\\
\multicolumn{4}{r}{\footnotesize$R^2_{\text{marginal}}=0.46$, $R^2_{\text{conditional}}=0.65$}\\
\bottomrule
\end{tabular}}
\end{table}

\begin{table}[htpb]
\centering
\caption{Mixed Effects Model for Self-Assessment with the users with the most and least data removed.}
\label{tab:LME-selfassessment-reducedpart}
\resizebox{\columnwidth}{!}{%
\begin{tabular}{lrrr}
\toprule
Variable & Coefficient  & Std. Err. & $p$-value \\
\midrule

Intercept & 0.128 & 0.055 & 0.020* \\
Avg. HR Difference & -0.016 & 0.005 & 0.001** \\
REM Sleep Duration Difference & 0.277 & 0.048 & <0.001*** \\
Bedtime End Hour Difference & 0.049 & 0.014 & <0.001*** \\
Bedtime Start Hour Difference & -0.048 & 0.013 & <0.001*** \\
$N$-Back Score Difference & 0.508 & 0.151 & <0.001*** \\
\multicolumn{4}{r}{\footnotesize * $p<0.05$, ** $p<0.01$, *** $p<0.001$}\\
\multicolumn{4}{r}{\footnotesize $R^2_{\text{marginal}}=0.15$, $R^2_{\text{conditional}}=0.18$}\\
\bottomrule
\end{tabular}}
\end{table}

To further validate this result, we model the dependency of self-reported sleep quality on the identified features with an ordinal cumulative link model and a Bayesian regression model, treating the self-reported sleep scores as ordinal values. The results can be seen in Table~\ref{tab:combined-ordinal-bayesian}.

\begin{table*}[htpb]
\centering
\caption{Ordinal Cumulative Link Model (left) and Bayesian Regression Model (right) for Self-Assessment.}
\label{tab:combined-ordinal-bayesian}
\begin{tabular}{lrrrrrrr}
\toprule
& \multicolumn{3}{c}{Ordinal Cumulative Link Model} & \multicolumn{4}{c}{Bayesian Regression Model} \\
\cmidrule(r){2-4} \cmidrule(l){5-8}
Variable & Coef. & Std. Err. & $p$-value & Coef. & Std. Err. & 5\% & 95\% \\
\midrule
Avg. HR Difference & $ -0.037$ & 0.011 & <0.001*** & $ -0.04$ & 0.01 & $-0.06$ & $-0.02$ \\
REM Sleep Duration Difference & 0.570 & 0.108 & <0.001*** & 0.57 & 0.11 & $0.35$ & $0.78$ \\
Bedtime End Hour Difference & 0.126 & 0.032 & <0.001*** & 0.13 & 0.03 & $0.07$ & $0.19$ \\
Bedtime Start Hour Difference & $ -0.096$ & 0.030 & 0.001** & $ -0.10$ & 0.03 & $-0.16$ & $-0.04$ \\
\textit{N}-Back Score Difference & 1.155 & 0.309 & <0.001*** & 1.16 & 0.32 & $0.54$ & $1.78$ \\
\bottomrule
\end{tabular}
\end{table*}

The ordinal model confirms the significance of the identified features and assigns them even larger effect sizes. Given that the Bayesian regression model aligns with the coefficients from the ordinal model, these larger effect sizes are likely more accurate than those obtained from the mixed effects model.

We also modeled \textit{N}-back scores using sleep features. Similar to the self-assessment model, we removed features to minimize VIF and AIC while maximizing $R^2$, resulting in a reduced feature set comprising absolute REM sleep duration, changes in REM sleep duration, and sleep self-assessment. The coefficients and significance values are detailed in Table~\ref{tab:LME-nback}, with all three remaining features showing high significance.

To assess whether the features identified in the self-assessment model consistently apply across all users, we divided participants into subgroups using a Gaussian mixture model based on all sleep features. We then applied the model from Table~\ref{tab:LME-selfassessment} to each subgroup and calculated Nakagawa and Schielzeth’s pseudo $R^2$~\cite{nakagawa_GeneralSimpleMethod_2013}. For the number of groups $n=2,3,4$, we found $R^2_{\text{marginal}}$ and $R^2_{\text{conditional}}$ values of $(0.157, 0.177)$, $(0.185, 0.208)$, and $(0.163, 0.190)$, respectively. For $n>4$, the mixed effects models did not converge, likely due to insufficient data per group.

The results of the subgroup mixed effects model coefficients are presented in Table~\ref{tab:LME-selfassessment-groups}. The explained variance and significance of specific parameters vary between groups. Group 1 (5 participants) shows the best fit, with significant effects for average HR difference, REM sleep duration difference, and bedtime end difference; bedtime start difference is marginally significant, while the \textit{N}-back score is not significant. Group 2 (11 participants) shows the second-best fit, with significant effects for bedtime end and start differences and \textit{N}-back score, while average HR difference and REM sleep duration difference are not significant. Group 3 (13 participants) shows the weakest fit, with only REM sleep duration difference being significant.

\begin{figure*}[htpb]
\centering
  \includegraphics[height=0.93\textheight]{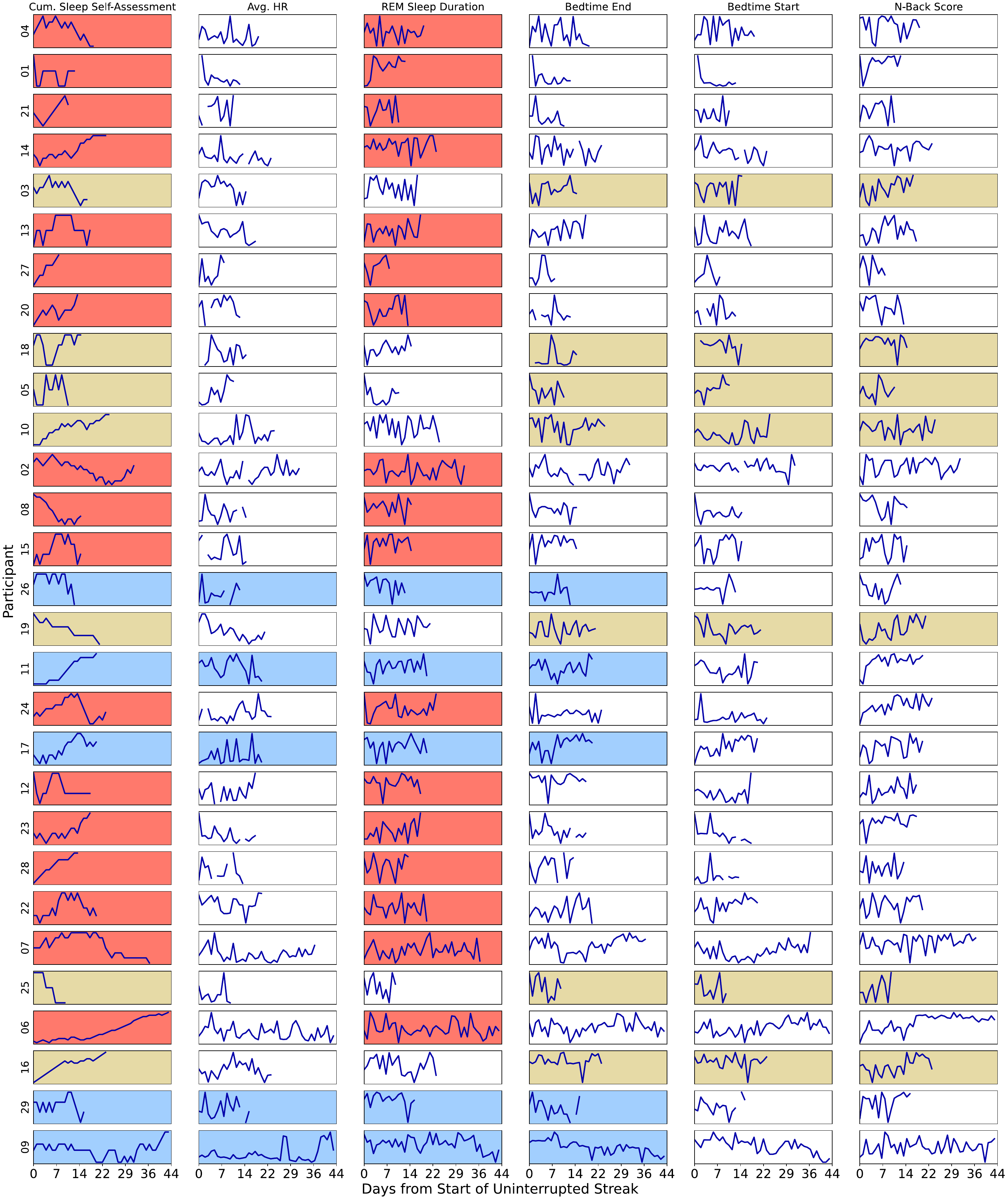}
  \caption{Cumulative self-reported sleep quality (first column) and the features included in the mixed effects models (columns 2–6) are shown for the longest consecutive data streak available for each participant. Groups 1, 2, and 3 are represented by blue, beige, and red, respectively. Dominant features from the group mixed effects models are highlighted with a filled background for each group.}
  \Description{Cumulative self-assessment (first column) and the features used in the mixed effects models (columns 2--6) per participant. The data displayed is for each participant's longest streak of available data. Blue, beige, and red indicate groups 1, 2, and 3, respectively. The dominant features identified from the group mixed effects models are marked by the filled background for every group.}
  \label{fig:time-series}
\end{figure*}

To visually verify whether self-reported sleep quality tracks the identified features, we generated time series plots of participants’ cumulative self-assessments for their longest uninterrupted data streaks. These plots, shown in Figure~\ref{fig:time-series}, display the five features identified in Table~\ref{tab:LME-selfassessment} and highlight each participant’s dominant feature based on group membership.

\begin{table}[htpb]
\centering
\caption{Mixed Effects Model for Self-Assessment for three identified groups.}
\label{tab:LME-selfassessment-groups}
\resizebox{\columnwidth}{!}{%
\begin{tabular}{lrrr}
\toprule
Variable & Coefficient  & Std. Err. & $p$-value \\
\midrule
Intercept & 0.110 & 0.072 & 0.124 \\
\textbf{Avg. HR Difference} & $\mathbf{-0.031}$ & $\mathbf{0.007}$ & \textbf{<0.001***} \\
\textbf{REM Sleep Duration Difference} & $\mathbf{0.479}$ & $\mathbf{0.114}$ & \textbf{<0.001***} \\
\textbf{Bedtime End Hour Difference} & $\mathbf{0.107}$ & $\mathbf{0.035}$ & \textbf{0.002**} \\
Bedtime Start Hour Difference & $ -0.056$ & 0.031 & 0.069 \\
\textit{N}-Back Score Difference & 0.348 & 0.409 & 0.395 \\
\multicolumn{4}{r}{Group 1: $R^2_{\text{marginal}}=0.36$, $R^2_{\text{conditional}}=0.36$}\\
\midrule
Intercept & 0.146 & 0.076 & 0.053 \\
Avg. HR Difference & $ -0.013$ & 0.010 & 0.176 \\
REM Sleep Duration Difference & 0.115 & 0.070 & 0.097 \\
\textbf{Bedtime End Hour Difference} & $\mathbf{0.084}$ & $\mathbf{0.021}$ & \textbf{<0.001***} \\
\textbf{Bedtime Start Hour Difference} & $\mathbf{-0.087}$ & $\mathbf{0.023}$ & \textbf{<0.001***} \\
\textbf{\textit{N}-Back Score Difference} & $\mathbf{0.785}$ & $\mathbf{0.214}$ & \textbf{<0.001***} \\
\multicolumn{4}{r}{Group 2: $R^2_{\text{marginal}}=0.20$, $R^2_{\text{conditional}}=0.25$}\\
\midrule
Intercept & 0.108 & 0.056 & 0.053 \\
Avg. HR Difference & 0.009 & 0.011 & 0.388 \\
\textbf{REM Sleep Duration Difference} & $\mathbf{0.358}$ & $\mathbf{0.074}$ & \textbf{<0.001***} \\
Bedtime End Hour Difference & 0.022 & 0.019 & 0.251 \\
Bedtime Start Hour Difference & $ -0.016$ & 0.018 & 0.367 \\
\textit{N}-Back Score Difference & 0.363 & 0.218 & 0.096 \\
\multicolumn{4}{r}{Group 3: $R^2_{\text{marginal}}=0.10$, $R^2_{\text{conditional}}=0.12$}\\
\midrule
\multicolumn{4}{r}{\footnotesize * $p<0.05$, ** $p<0.01$, *** $p<0.001$}\\
\multicolumn{4}{r}{Overall: $R^2_{\text{marginal}}=0.185$, $R^2_{\text{conditional}}=0.208$}\\
\bottomrule
\end{tabular}}
\end{table}

The ordinal model estimates the coefficients generally higher than the mixed effects model using continuous variables.

\section{Discussion}


The correlation analysis in Section~\ref{sec:corr} reveals a very weak tendency for \textit{N}-back scores to increase with a higher number of restless periods and later wake-up times. While we can speculate some causes for the later wake-up times—such as longer sleep duration or a less stressful, more relaxed day—explaining why more restless periods might improve working memory performance is more challenging. Ultimately, the correlation is negligible. Surprisingly, there is no strong correlation with total sleep duration, any specific sleep phase duration, or sleep efficiency, as previous work might suggest \cite{kuriyama_SleepAcceleratesImprovement_2008}

The coefficients from the mixed effects model indicate that REM sleep duration has the most substantial effect on predicting \textit{N}-back scores, which reflect working memory capacity. Furthermore, changes in REM sleep duration exert a significant negative impact; if the absolute REM sleep duration for the current night is kept constant, a longer REM sleep duration on the previous night results in a similar effect. This suggests that the impact of REM sleep duration accumulates over multiple nights. Additionally, the participant’s sleep self-assessment emerges as a key factor. Assuming the \textit{N}-back score reliably measures momentary cognitive readiness, it appears that participants have insights into their own sleep quality that extend beyond what the ring can measure.

\subsection{Sleep Self-Reports}

As described in Section~\ref{sec:corr}, self-reports of sleep generally correlate weakly and positively with Oura's reported sleep score, sleep-phase durations, percentage or REM sleep, and sleep efficiency and longer bedtimes. This is expected and confirms that participants' self-reports are at least partially based on the general parameters of their sleep. No significant correlation could be found for other known markers of sleep quality, such as HR or HRV. While the general direction of coefficients confirms previous studies (e.g. 
\cite{schlagintweit_EffectsSleepFragmentation_2023, zmijewska_AssociationHeartRate_2024,cretikos_RespiratoryRateNeglected_2008}) about sleep quality -- lower HR, higher HRV, and lower BR indicate better sleep -- the effect sizes are too small to be significant. The negative correlation with deep sleep percentage can partially be explained through the zero-sum characteristic of sleep phases. If the REM sleep percentage increases, the deep and/or light sleep percentage will decrease. The fact that light sleep percentage shows a significant correlation to higher self-assessment of sleep could nonetheless indicate that deep sleep could have some negative causal relationship to self-reported sleep, possibly due to increased sleep inertia after waking up~\cite{trotti_WakingHardestThing_2017}, but further research is necessary.

Concerning the mixed effects, ordinal, and Bayesian models, the features identified mostly confirm the previous correlation analysis, with REM sleep duration and longer bedtimes being predictors of self-reported sleep. Here, we find that self-reported sleep improves with lower average overnight HR, a marker of sleep quality as identified in previous work~\cite{sajjadieh_AssociationSleepDuration_2020}. \textit{N}-back scores also show significant predictive qualities for self-reported sleep quality. Together with the previous finding that self-assessment is a significant predictor of \textit{N}-back scores, this could indicate that participants self-reports correctly identify their sleep quality and thereby improved working-memory performance.

By assigning the participants into subgroups, we find that participants' self-reports respond to the individual features to a varying degree. This is a novelty all by itself, since it shows that the alignment of perceived and measured sleep quality is not equal among participants and varies in magnitude (coefficients) and quality (features). We identified three groups: For the first group, consisting of five participants,  average HR difference, REM sleep duration difference, and the difference in wake-up time have the strongest effect, with $R^2=0.36$ for fixed effects being the highest among the groups. The second group, consisting of 11 participants, showed the highest response to changes in wake-up time, sleep time, and \textit{N}-back score and the second-best model fit. For group 3, consisting of 13 participants, the difference in REM sleep duration is the strongest predictor.

Irrespective of the subgroup, weak alignment of self-assessment with sleep markers indicates that participants can gain information that is not perfectly aligned with their feelings about their sleep from the sensor readings. If we consider the different response rates of sub-groups to sleep markers, we can conclude that the level of information gain is not equal among participants. Another indication of information gain is that sleep self-reports are a significant predictor of working memory performance, in addition to REM sleep duration. This means the model has higher predictive quality when self-reports are combined with physiological measures and that the information therein is at least partially supplemental to the physiological measures themselves, providing an information gain. The varying model fit indicates that this information gain varies across the subgroups. If this wasn't the case, sensor readings alone should be sufficient to model the momentary working memory capacity.

While we do not dive deep into what differentiates the sub-groups besides their sensitivity to different sleep markers, the fact that one group is influenced by the end of their bedtime but not by the start could indicate chronotypes, although further research is needed. An in-depth analysis of identified groups is out of the scope of this work, however, and would benefit greatly from a larger participant base, since there are indications that the currently identified number of 3 groups comes from data limitations.

Visual inspection of the cumulative self-assessment and corresponding sleep features mostly identifies no real patterns or tracking, but since $R^2$ of individual group models ranges between 0.12 and 0.36, and self-assessment is influenced by combinations of values, this is to be expected.

\subsection{Implications for HCI Research and Design}

The HCI community has extensively been researching self-tracking technologies, whether for monitoring sleep, assessing progress in the gym, or tracking food consumption.
A common observation in the real-world use of these technologies is the eventual drop in user interest, with data collection often falling off rapidly after a relatively short period of time~\cite{Gouveia2015Activitytrackers}. This indicates that these devices often fail to provide sufficient value to end-users.

In addition to these general challenges, the tracking of sleep quality, in particular, is complicated by users being largely unconscious during the data collection, unlike many other self-tracking activities, during which users are generally more aware of the activity being performed and its characteristics (i.e., running a certain distance).
This inherent disconnect can also result in uncertainty towards the reliability and applicability of sleep data~\cite{Kuosmanen2022Sleeptracking, Ravichandran2017SenseSleep}, potentially further reducing interest in collecting sleep data.
Prior work highlights that users differ in their prioritization of different sleep features when assessing their sleep quality~\cite{Liang2016SleepExplorer}.
For example, Kuosmanen et al.\ found that close to 80\% of participants in a longitudinal sleep self-tracking study focused on metrics related to sleep stages in their assessment of sleep quality~\cite{Kuosmanen2022Sleeptracking}.
However, as we are unable to control the number of hours spent in a particular sleep stage~\cite{Ravichandran2017SenseSleep}, this metric is not particularly helpful in driving changes in sleep quality.

Our results suggest that participants' self-reported sleep quality correlates with the data collected through a self-tracking device.
Assessing the literature on sleep self-tracking, however, the usefulness of these metrics can be questioned given their limited direct applicability.
Consequently, based on our quantitative analysis and the insights from prior work on users' needs and challenges in sleep tracking~\cite{Kuosmanen2022Sleeptracking, Ravichandran2017SenseSleep}, we suggest HCI researchers investigate the possibility of incorporating metrics that are more immediately relevant to end-users. Here, it is valuable to note that our results on the \textit{N}-back scores highlight significant predictive power for self-reported sleep quality.
Providing users with insights into their real-world cognitive readiness---based on their quantitative sleep data---could, therefore, be an example of such an actionable insight that might support end-users in, for example, organizing their daily routine.


We argue that this approach does not only have the potential to enhance engagement but also supports the design of data delivery systems that promote better user literacy. Presenting actionable insights—such as the relationship between sleep quality and changes in real-world cognitive readiness—can help users make sense of rather abstract data and relate it to their daily goals or behaviours~\cite{Li_2010}. Prior studies suggest that aligning data representation with user priorities and cognitive models fosters a deeper understanding and sustained use of self-tracking technologies~\cite{Epstein_2015, Rapp2016}. Furthermore, visualizing connections between sleep metrics and their real-world impacts has been shown to improve user trust and motivation to track their progress~\cite{Choe_2014, Swan2009}.

By making data more accessible and actionable, e.g., by providing insights, such as the relationship between sleep quality and the ability to do better or worse on a task due, users can better understand and apply abstract data to their daily lives. Aligning data with user priorities and cognitive models increases comprehension, as users are more likely to engage with relevant information~\cite{Li_2010, Epstein_2015}. Moreover, visualizing connections between sleep metrics and real-world outcomes further aids in understanding and improving trust and motivation~\cite{Swan2009, Choe_2014}. Ultimately, this system empowers users to make informed decisions, fostering data literacy by helping them interpret and act on information effectively~\cite{Rapp2016, Li_2010}.


\subsection{Limitations}

While the findings are promising, the complex interactions between sleep quality, self-reports and working memory performance are a strong limiting factor. We have found correlations and predictive qualities between self-reports and sleep metrics, but it is unclear what or if there is a causal relation.

One limitation is the possibility of self-bias: The self-reported sleep quality in the first part of the experiment application could have biasing effects on the working memory test. For example, participants may no longer perform the \textit{N}-back test with full diligence if they reported having slept poorly or significantly worse. Conversely, effects like behavioral confirmation~\cite{snyder_WhyDoesBehavioral_1995} or suggestion~\cite{michael_SuggestionCognitionBehavior_2012} could lead to better scores after higher reported sleep quality. This would be a possible explanation for the mutual predictive qualities of both values. Another limiting factor could be the language of the application chosen by the participants. It is possible that participants understand nuances of the question about their sleep differently depending on language and, therefore, weigh different characteristics differently.
A single-item sleep questionnaire may not capture the same level of information as a multiple-item measure (though the latter may increase the burden on participants).
Bilingualism may have effects on working memory \cite{calvo_ImpactBilingualismWorking_2016, anton_ImpactBilingualismExecutive_2019, yang_BilingualsWorkingMemory_2017}, although the magnitude and direction of effect are still unclear. Since most students in our study are bilingual, there may be more potential to investigate in this direction, especially considering the language diversity of the participants.
While we asked participants how long it took the to get used to the N-back test, there is still a weak but significant correlation between test number and N-back score after discarding those days ($\rho_{\textit
pearson}=0.13$, $p<0.001$), i.e. there is a slight learning effect even after participants are comfortable with the test.
A further limiting factor is that due to gaps in experiment participation, the longest possible streak for each participant is mostly much shorter than the total number of days the participant provided. Only one participant missed no days. Due to the relative nature of self-reports, the remaining cumulative self-assessment after a break is offset by an unknown amount during the break. To circumvent this, a much stricter experiment setup is needed, where participants must self-report daily. The small participant base is an additional limiting factor. More participants and a larger dataset could enable other analysis methods, e.g., (deep) learning-based methods.

\section{Conclusion}

In this study, we asked 29 participants to rate their sleep daily compared to the previous night's sleep. We asked them to wear Oura rings to get sensor readings about specific sleep markers. Participants also conducted short working memory tests after self-reporting their sleep. The whole experiment procedure took place in-the-wild, meaning in their usual environments while following their usual daily and nightly (sleep) schedules.
We found weak associations between self-reported sleep quality, sleep durations, sleep efficiency, and bedtimes. We found weak associations between the waking up time and restless periods for working memory scores.
We identified the differences in REM sleep duration, average overnight HR, working-memory scores, and bedtimes as being most predictive of sleep quality self-reports. For working memory scores, we found the REM sleep durations of the two previous nights and self-assessments to be most predictive. From these results, we conclude that sensor readings and sleep features represent an information gain for the wearers of the device since sensor readings do not fully capture the variance in self-reports. Still, self-reports enhance sensor readings in predicting working memory scores.
We identified three subgroups with different sensitivities toward the identified sleep markers. From this, we conclude that the alignment of self-reports to sleep markers and information gained from the sleep trackers is not homogeneous among the participant base.
\begin{acks}
This work was supported in part by grants from JST ASPIRE (JPMJAP2403), JSPS Grant-in-Aid for Scientific Research (23KK0188) and the Grand challenge of the Initiative for Life Design Innovation (iLDi).
\end{acks}

\bibliographystyle{ACM-Reference-Format}
\bibliography{software, zotero, prevbib, pastbib2, CHI25}

\end{document}